\documentclass[prapplied,rsi, amsmath,amssymb,reprint,superscriptaddress]{revtex4-1}
 
\usepackage{longtable}
\usepackage{graphicx}
\usepackage{color}
\usepackage{soul}

\usepackage[version=3]{mhchem} % Formula subscripts using \ce{}
\usepackage{blindtext}
\usepackage{bm}
\usepackage{array}

\usepackage{multirow}

\newcolumntype{C}{>{\centering\arraybackslash}p{4em}}
\newcolumntype{D}{>{\centering\arraybackslash}p{6em}}

\usepackage{longtable}

\usepackage{array}
\newcolumntype{L}[1]{>{\raggedright\let\newline\\\arraybackslash\hspace{0pt}}m{#1}}
\newcolumntype{C}[1]{>{\centering\let\newline\\\arraybackslash\hspace{0pt}}m{#1}}
\newcolumntype{R}[1]{>{\raggedleft\let\newline\\\arraybackslash\hspace{0pt}}m{#1}}

\begin{document}

% \title{Crystal nucleation in nanoparticles: \\ How to extract critical nucleus on top of the free energy barrier?}

%\title{Crystallization processes of nanoparticles: a structural analysis of transition state ensembles}
\title{Harvesting nucleating structures in nanoparticle crystallization: \\ The example of gold, silver and iron}

\author{Arthur France-Lanord}
\email{arthur.france-lanord@cnrs.fr}
\affiliation{Sorbonne Université, Muséum National d’Histoire Naturelle, UMR CNRS 7590, Institut de Minéralogie, de Physique des Matériaux et de Cosmochimie, IMPMC, F-75005 Paris, France}

\author{Sarath Menon}
\affiliation{Max-Planck-Institut f\"ur Eisenforschung GmbH, D-40237 Düsseldorf, Germany}

\author{Julien Lam}
\email{julien.lam@cnrs.fr}	
\affiliation{Univ. Lille, CNRS, INRA, ENSCL, UMR 8207, UMET, Unité Matériaux et Transformations, F 59000 Lille, France}
\affiliation{Centre d’élaboration des Matériaux et d’Etudes Structurales, CNRS (UPR 8011), 29 rue Jeanne Marvig, 31055 Toulouse Cedex 4, France}

\begin{abstract}

The thermodynamics and kinetics of nanoparticle crystallization, as opposed to bulk phases, may be influenced by surface and size effects. We investigate the importance of such factors in the crystallization process of gold, silver, and iron nanodroplets using numerical simulations in the form of molecular dynamics combined with path sampling. This modeling strategy is targeted at obtaining representative ensembles of structures located at the transition state of the crystallization process. A structural analysis of the transition state ensembles reveals that both the average size and location of the critical nucleation cluster are influenced by surface and nanoscale size effets. Furthermore, we also show that transition state structures in smaller nanodroplets exhibit a more ordered liquid phase, and differentiating between a well-ordered critical cluster and its surrounding disordered liquid phase becomes less evident. All in all, these findings demonstrate that crystallization mechanisms in nanoparticles go beyond the assumptions of classical nucleation theory. 
\end{abstract}

\maketitle

\section{Introduction}

The formation of nanocrystals plays a crucial role in numerous research fields including material science\cite{Rathinavel2021Jun,Farzin2021Mar,Baig2021Mar}, geology\cite{Hochella2008Mar,Schafer2012Feb}, environmental science\cite{Ronkko2017Jul,Kulmala2014Apr,Bundschuh2018Dec} and biology\cite{Mathivanan2021Jun,Prasad2014Dec}. Beyond these applied interests, crystal formation transposed to nanoscale systems exhibits many additional challenges when compared to its bulk analogue. Indeed, the preponderance of surface and the natural occurrence of finite-size effects drastically increase the complexity of nucleation mechanisms\cite{Lee2016Jun,Thanh2014Aug,Jun2022Apr,Zhang2012Mar}. As such, classical nucleation theory -- which was widely employed to understand crystal formation in bulk phases --, was proven inadequate to describe many experimental observations of the nanocrystal nucleation, including the competition between several polymorphic phases\cite{Lam2018Mar,Amodeo2020Oct,Wells2015Apr,Tarrat2023Apr}, and the emergence of pre-nucleation clusters\cite{Ramamoorthy2020Aug,Ramamoorthy2020Aug,Schiener2015Jun}.

Along with experimental measurements, numerical simulations have been pivotal in providing further insights on the inherent mechanisms\cite{Schoonen2022Dec,Lutsko2019Apr}, including atomistic observations based on molecular dynamics\cite{Quigley2011Jan,Pavan2015Nov,Lam2018Mar,Amodeo2020Oct}. In this research field, which consists in observing \textit{in silico} the birth of crystals\cite{Sosso2016Jun}, the necessity to overcome free energy barriers is the major challenge, be it for bulk phases or nanoscale systems. In this context, while rare-event sampling methods have been extensively developed to help characterizing nucleation processes\cite{Giberti2015Mar,Pietrucci2017Nov,Bolhuis2002Oct,Bolhuis2021Apr}, their application to the formation of nanocrystals have so far been limited to very few examples\cite{Quigley2011Jan,Pavan2015Nov,Amodeo2020Oct}. Although highly informative, these studies focused on enhanced sampling (in the form of metadynamics\cite{Laio2002escaping}) to overcome free energy barriers, which requires biasing the dynamics along a specified collective variable. This bias influences the dynamics as well as the sampled configurations, in a possibly unrealistic fashion, depending on many factors, including the quality of the collective variable\cite{bussi2020using}. Recently, several groups have proposed to use biasing techniques along with trajectory-based sampling methods; this allows eliminating the influence of biaising on the investigated kinetics \cite{Finney2022Jul,Borrero2016Oct,Falkner2024Mar}. 

In this work, we employ an alternative simulation strategy which consists in combining metadynamics along with committor analysis\cite{peters2006using} and transition path sampling\cite{dellago2002transition} in the form of aimless shooting\cite{Peters2006aimless,Peters2007extensions}. Metadynamics is only used to formally identify, through committor analysis, structures close to or belonging to the transition state ensemble. These are then used as starting points to sample unbiased reactive trajectories and transition state configurations, using aimless shooting. We therefore obtain ensembles of critical structures located on top of the free energy barrier, without the need for biasing along any collective variable. We apply this methodology to investigate crystal formation in nanodroplets composed of three different metals, namely gold, silver, and iron, which are currently considered in many technological applications\cite{Zhang2020,Yaqoob2020May,Ebrahiminezhad2018Feb}. Upon decreasing the droplet size, we demonstrate that the critical nucleus gets smaller, and tends to move towards the droplet's surface. In addition, as observed in bulk crystallization and denoted as pre-nucleation mechanisms, the surrounding liquid becomes more ordered when compared to a normal liquid. Finally, the interface between the critical nucleus and its surroundings gets more diffuse. Altogether, our results strongly suggest the existence of crystallization processes beyond classical nucleation theory. 

\section{Methodology}\label{s:methods}

We study three different materials, namely gold, silver, and iron. Their interactions are modeled using three different EAM interatomic potentials\cite{Mendelev2003, Williams2006May,Chamati2004Nov} chosen for their good match with experimental data regarding both bulk and surface properties\cite{Combettes2020Sep}. Throughout this work, molecular dynamics (MD) simulations are carried out using the LAMMPS package (v.\,23 Jun 2022)\cite{thompson2022lammps}, supplemented by PLUMED (v.\,2.7.5)\cite{bonomi2019promoting,tribello2014plumed} for enhanced sampling as well as for the definition of the collective variables. The time step is set to $1$\,fs and a Nosé-Hoover chain of three thermostats is applied with a damping factor of $100$\,fs. To measure crystal ordering, the averaged Steinhardt parameter ($\bar{q_6}$) is computed with a cutoff equal to 3.5\,\AA\, which roughly corresponds to the end of the first peak of the radial distribution in the liquid regime. We then use the $\bar{q_6}$ value of 0.35 as a threshold discriminating liquid and crystal structures\,\cite{lechner2008accurate}. During a simulation, more than one crystalline cluster can appear simultaneously. Therefore, we define $N_{\text{crys}}$ as the size of the largest cluster with a cutoff parameter equal to 4\,\AA using the Cluster analysis as implemented in Ovito (v. 3.7.7)\cite{Stukowski2009Dec}. Finally, we consider four different nanoparticles made of 500, 1000, 1500 and 2000\, atoms.

With our current simulation protocol, we purposely assume that the atomic precursors concerned by the nanoparticle synthesis coalesce at some point to form our initial liquid nanodroplets. Yet, depending on the thermodynamic conditions, it is also possible that nanocrystals form through atom-by-atom growth directly in the ordered phase without going through the liquid phase. Few works attempted at modeling the whole gas to nanocrystal process which required the use of unrealistically fast cooling rates\cite{Forster2019Oct,Zhao2015Jan,Kesala2007May,Rossi2023Jun}. Meanwhile, authors also chose to deposit atoms on an already formed solid seed, while varying the deposition rate\cite{Goniakowski2010Apr,Wells2015Apr,Xia2021May}. This therefore involves investigating a different process than in the present work, namely a deposition (or desublimation) transition. The methodology we propose here, consisting in combining metadynamics with committor analysis and aimless shooting simulations, can of course still be employed in this alternative situation. 

\section{Results}

\subsection{Measuring the melting temperature}

\begin{figure}[h!]
    \centering
    \includegraphics[width=8.6cm]{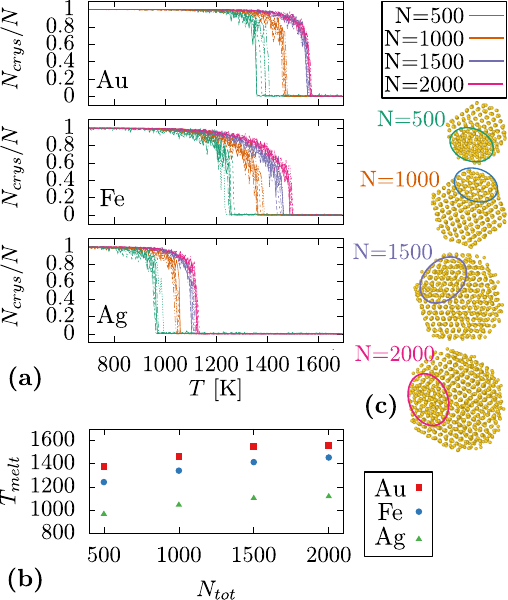}
    \caption{Measurement of the melting temperature as a function of the droplet size. (a) Crystal proportion as a function of the temperature during five different heating rates shown in different dash types. (b) Extracted melting temperature averaged over the five different heating rates. (c) Snapshot corresponding to the onset of melting shown for Au nano-crystal.}
    \label{Tmelt}
\end{figure}

Our goal is to sample many configurations at the transition state ensemble of nanoparticle crystallization. In order to compare different sizes as well as different materials, we position ourselves at the same degree of supercooling, which requires measuring the melting temperature. To this end, we start by generating crystalline configurations by carving spheres from a bulk crystal that are then structurally optimized at 0\,K for each materials and each nanoparticle sizes.  We start simulations in the canonical ($nVT$) ensemble at 700\,K, well below the melting temperature, and progressively ramp up the temperature up to 1700\,K. Five different simulation lengths are considered: 10, 9, 8, 7, 6 and 5\,ns which correspond to a range in heating ramp going from 100\,K/ns to 200\,K/ns. From the recorded trajectories, we measure the temperature evolution of $N_{\text{crys}}$ [See Fig.\,\ref{Tmelt}.a] and observe that the heating ramp does not significantly influence the results. We then measure a melting temperature by fitting the obtained curves with a hyperbolic tangent function and extracting a critical temperature, for which corresponding snapshots are displayed in Fig.\,\ref{Tmelt}.c. Fig.\,\ref{Tmelt}.b presents the melting temperatures averaged over the five considered heating ramp, and confirms that the melting temperature is negatively correlated with the nanoparticle size, as observed in experiments\cite{Lai1996Jul}. We note that the obtained melting temperatures is highly dependent on some arbitrary choices including $N_{\text{crys}}$ as the collective variable, the hyperbolic tangent fit, and the employed empirical interatomic potential. As such, the obtained values for $T_{\text{melt}}$ should not really be compared with experimental results although our approach remains very similar to recent attempts at measuring melting temperatures in nanoparticles\cite{Zeni2021Oct,Delgado-Callico2021,Rossi2018Jun}. What is important here is to use a consistent method for all materials and droplet sizes, in order to control the degree of supercooling employed in the remaining part of the study. Since we work at different temperatures for different droplet sizes, it is important to disentangle the effects of both factors. For this reason, we have performed additional simulations on the largest droplet, at the temperature corresponding to the smallest droplet. In the Supplementary Material, we show that temperature alone does not explain the results we observe over different droplet sizes. 

%the crystalline proportion of samples by computing, for each atom, the averaged sixth-order Steinhardt parameter ($\bar{q_6}$). Atoms showing $q_6 > \afl{0.25?}$ are labeled as being crystalline. The melting temperature is extracted by \afl{?}. The crystalline proportion as a function of applied temperature and nanoparticle size is reported in Fig. \ref{fig:rad}. As can be seen, the heating rate does not affect to a large extent the temperature evolution of the crystalline fraction. Although overestimated (the experimental melting temperature of bulk gold is 1336~K\cite{Rumble2017crc}), the measured melting temperatures are indeed anticorrelated with particle size. 

\subsection{Obtaining a first transition state configuration}

\begin{figure*}[ht]
    \centering
    \includegraphics[width=16cm]{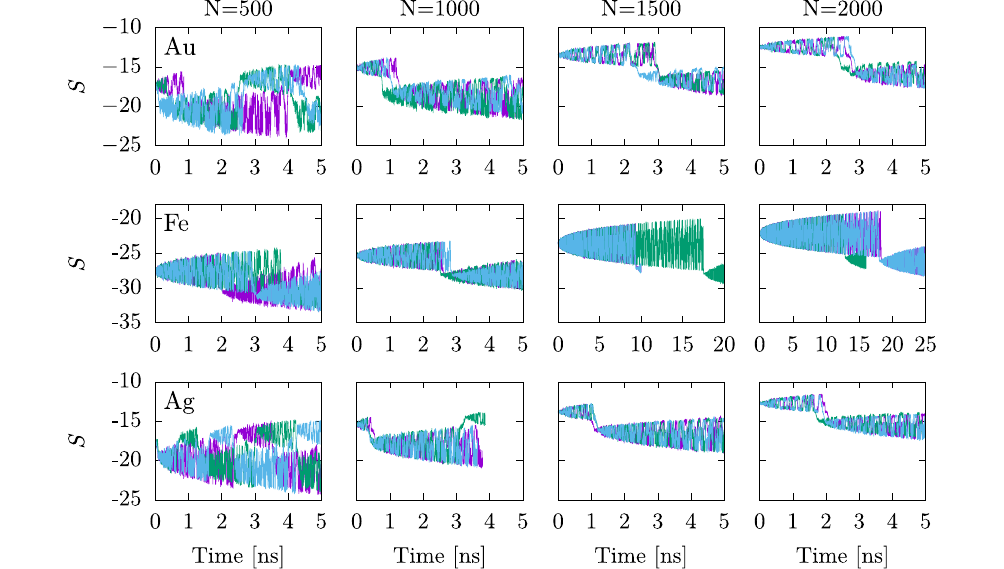}
    \caption{Metadynamics results for different droplet sizes and materials. In each case, the three different curves correspond to three different initial positions/velocities, with the remaining simulation parameters are kept constants.}
    \label{Meta}
\end{figure*}

In the following, all simulations are carried out at $T=0.85\,T_{\text{melt}}$, for all nanoparticle sizes and materials. In order to find a first transition state connecting the liquid and crystalline states, we perform metadynamics simulations with fixed Gaussian heights \cite{Laio2002escaping,bussi2020using,Laio2008Nov}. Simulations are initialized in the liquid regime obtained from the previously melted simulations and thermalized at the chosen temperature during $100$\,ps. Then, the metadynamics bias is chosen along the effective entropy $S$, derived from liquid state theory: 
\begin{equation}\label{eq:s}
S = - 2 \pi \rho \int_0^{+\infty} \left[ g(r) \ln g(r) - g(r) + 1 \right] r^2 \mathrm{d}r, 
\end{equation}
where $\rho$ is the atomic density measured in the whole simulation box and $g(r)$ is the radial distribution function. Every 1\,ps, we add bias in the form of Gaussian functions of width and height respectively equal to 0.5\,$k_B$ and 1\,kJ/mol for all sizes and materials. In all cases, we perform three independent simulations using different initial configurations and atomic velocities. Results reported in Fig. \ref{Meta} show that $S$ allows for overcoming the free energy barrier of nucleation, which is consistent with previous works employing the same collective variable\cite{Piaggi2017Jul,Piaggi2017Sep,Amodeo2020Oct,Lam2023Jan}. More importantly, it appears that larger droplets seem to require longer simulations in order to reach a first transition. However, it does not necessarily mean that the associated free energy barrier is larger, since a proper measurement of the free energy barrier requires overcoming numerous technical challenges especially related to the definition of a representative collective variable.\cite{Lam2023Jan} Since this is not the scope of this work, we will not use metadynamics calculations to estimate a free energy barrier, but instead only to obtain a first transition state configuration. In particular, the committor probability $P_{\text{crys}}$, \textit{i.e.} defined as the probability that a configuration $\mathbf{X}$ will reach the crystalline state before reaching the liquid state, is measured for configurations extracted from the metaydnamics simulations. For that purpose, we initialize unbiased molecular dynamics simulations with atomic velocity drawn from the Maxwell-Boltzmann distribution, and monitor the committment of the system towards the liquid or the crystalline metastable states. Both basins are defined using $\overline{q_6}$ averaged over all of the atoms. Such a choice of order parameter rather than $N_{\text{crys}}$ for instance allows us to use an intensive property of the system that does not depend on the droplet size $N$. For all the materials, the crystalline basin corresponds to  $\overline{q_6} > 0.35$. Since iron liquid is inherently less ordered than the two noble metals, two different values are chosen for the liquid basin of gold and silver ($\overline{q_6} < 0.225$) and iron ($\overline{q_6} < 0.175$). At this stage, we are solely interested in coarsely characterizing transition states: we therefore only used 10 trajectories to measure $P_{\text{crys}}$, and defined a first transition state as matching $P_{\text{crys}} \in [0.3, 0.7]$. From each of the three metadynamics simulations performed for all of the considered materials and droplet sizes, we managed to extract one transition state configuration and several unbiased trajectories reaching either the crystal or the liquid regime.

%\afl{Pour générer un chemin réactif initial pour l'aimless shooting, il faut une trajectoire qui connecte les deux bassins. Tu pourras élaborer?}

\subsection{Sampling many transition state configurations and reactive paths}\label{sub:as}

Aimless shooting\cite{Peters2006aimless,Peters2007extensions} is a transition path sampling method which does not require biasing following a particular collective variable. The principle is the following: a configuration $\mathbf{X}(t)$, which is part of an initial path, is propagated backward and forward in time (initial velocities are drawn from the Maxwell-Boltzmann distribution). The dynamics are interrupted once either the liquid ($\overline{q_6} < 0.175$) or the crystalline state ($\overline{q_6} > 0.40$) is reached, with basin definitions slightly more stringent than for the committor estimation. If the new path connects both basins, it is accepted and used for the next iteration. If it connects the same basin, it is rejected; in this case, the former path is used for the next iteration. Then, either configuration $\mathbf{X}(t)$, $\mathbf{X}(t + \Delta t)$, or $\mathbf{X}(t - \Delta t)$ -- \textit{i.e.} configurations separated by a time step $\Delta t$ before or after $\mathbf{X}(t)$ in the current path -- is selected as a new shooting point. This corresponds to the original three-point aimless shooting algorithm\cite{Peters2006aimless}. Iterating allows to sample transition pathways, which, since they obviously pass through the separatrix, leads also to sampling configurations from the transition state ensemble. The acceptance criterion acts as an effective restoring force towards the separatrix: biaising is performed in trajectory space instead of configuration space. Selecting the right value for $\Delta t$ is not trivial. A too small time step would lead to sampling highly correlated configurations, while a too large value would lead to a low path acceptance ratio. We set $\Delta t$ equal to 2\,ps, which leads to a global average acceptance ratio of 28.6\,\%. For each of the three initial paths, we perform three independent aimless shooting simulations with randomized initial velocities. In total, we sample 47163 pathways, out of which 13358 connect both basins. This amounts to a total sampling time slightly larger than 3.3 $\mu$s. System-dependent detailed statistics are reported in Table \ref{t:aimless}. The average accepted trajectory length is 70.6 ps, with an associated standard deviation of 56.1 ps. Trajectory length distributions for each of the considered nanoparticle are reported in the Supplementary Material. 

\begin{table}[]
\centering
\caption{Statistics of the aimless shooting simulations for each system: number of iterations and number of accepted configurations, with bold numbers being sums over all nine independent runs (three initial paths, three velocity initializations). }
\label{t:aimless}
\begin{tabular}{ C{.5cm} C{1cm} C{.8cm} C{.8cm} C{.8cm} C{.8cm} C{.8cm} C{.8cm} }
\multicolumn{2}{c}{\multirow{2}{*}{System}} & \multicolumn{3}{C{2.4cm}}{$N_{\text{iterations} (\Delta t)}$} & \multicolumn{3}{C{2.4cm}}{$N_{\text{accepted}}$} \\

 & & avg & min & max & avg & min & max \\
\hline\hline
\multirow{8}{*}{Au} & \multirow{2}{*}{500} & 453 & 258 & 822 & 127 & 87 & 172 \\
 & & \multicolumn{3}{c}{\textbf{4073}} & \multicolumn{3}{c}{\textbf{1141}} \\\cline{2-8}
 & \multirow{2}{*}{1000} & 350 & 236 & 479 & 118 & 61 & 182 \\
 & & \multicolumn{3}{c}{\textbf{3146}} & \multicolumn{3}{c}{\textbf{1063}} \\\cline{2-8}
 & \multirow{2}{*}{1500} & 298 & 221 & 384 & 79 & 58 & 102 \\
 & & \multicolumn{3}{c}{\textbf{2678}} & \multicolumn{3}{c}{\textbf{711}} \\\cline{2-8}
 & \multirow{2}{*}{2000} & 231 & 198 & 286 & 82 & 62 & 118 \\
 & & \multicolumn{3}{c}{\textbf{2077}} & \multicolumn{3}{c}{\textbf{738}} \\

% Still need to fill Ag
\hline\hline
\multirow{8}{*}{Ag} & \multirow{2}{*}{500} & 668 & 447 & 1062 & 187 & 21 & 336 \\
 & & \multicolumn{3}{c}{\textbf{6015}} & \multicolumn{3}{c}{\textbf{1683}} \\\cline{2-8}
 & \multirow{2}{*}{1000} & 534 & 424 & 875 & 153 & 57 & 221 \\
 & & \multicolumn{3}{c}{\textbf{4807}} & \multicolumn{3}{c}{\textbf{1374}} \\\cline{2-8}
 & \multirow{2}{*}{1500} & 479 & 371 & 683 & 134 & 61 & 204 \\
 & & \multicolumn{3}{c}{\textbf{4312}} & \multicolumn{3}{c}{\textbf{1204}} \\\cline{2-8}
 & \multirow{2}{*}{2000} & 391 & 296 & 586 & 117 & 49 & 169 \\
 & & \multicolumn{3}{c}{\textbf{3518}} & \multicolumn{3}{c}{\textbf{1054}} \\

% Still need to fill Fe
\hline\hline
\multirow{8}{*}{Fe} & \multirow{2}{*}{500} & 561 & 404 & 837 & 153 & 103 & 215 \\
 & & \multicolumn{3}{c}{\textbf{5053}} & \multicolumn{3}{c}{\textbf{1379}} \\\cline{2-8}
 & \multirow{2}{*}{1000} & 582 & 525 & 653 & 148 & 101 & 182 \\
 & & \multicolumn{3}{c}{\textbf{5240}} & \multicolumn{3}{c}{\textbf{1333}} \\\cline{2-8}
 & \multirow{2}{*}{1500} & 394 & 330 & 431 & 99 & 29 & 141 \\
 & & \multicolumn{3}{c}{\textbf{3546}} & \multicolumn{3}{c}{\textbf{890}} \\\cline{2-8}
 & \multirow{2}{*}{2000} & 300 & 215 & 364 & 85 & 64 & 107 \\
 & & \multicolumn{3}{c}{\textbf{2698}} & \multicolumn{3}{c}{\textbf{766}} \\
\end{tabular}
\end{table}

\begin{figure*}[ht]
    \centering
    \includegraphics[width=\linewidth]{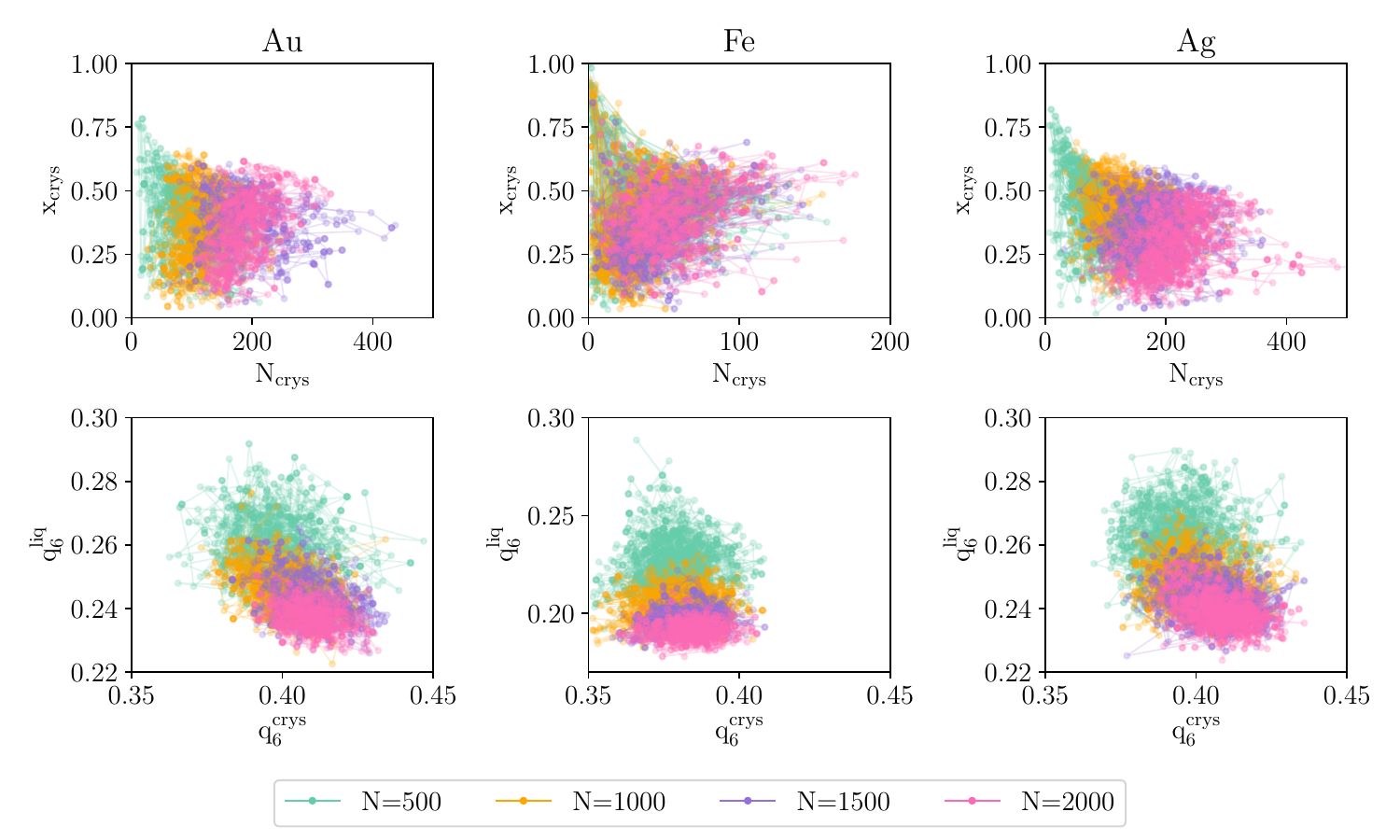}
    \caption{Trajectories of configurations accepted by aimless shooting in both $(N_{\text{crys}}, x_{\text{crys}})$ and $(q_6^{\text{crys}}, q_6^{\text{liq}})$ spaces. }
    \label{fig:astrj}
\end{figure*}

In Fig. \ref{fig:astrj}, we are reporting trajectories of configurations accepted by aimless shooting in two two-dimensional collective variable spaces: $(N_{\text{crys}}, x_{\text{crys}})$ and $(q_6^{\text{crys}}, q_6^{\text{liq}})$. For $N_{\text{crys}}$, we use the definition already introduced in section \ref{s:methods}. $x_{\text{crys}}$ corresponds to the ratio between the radial position of the mass barycenter of the largest crystalline cluster and the radius of the droplet. Finally, we defined $q_6^{\text{crys}}$ and $q_6^{\text{liq}}$ as the averaged values of $\overline{q_6}$ when selecting respectively only the atoms in the largest crystalline cluster, and the remaining ones. These are descriptors of the structuration of the critical nucleus and its surroundings. Although we are initializing independent aimless shooting simulations with three different pathways, as well as three different initial velocities, all simulations seem to converge to the same portions of configuration space -- projected along these collective variables. This seems to indicate that our sampling of the transition state ensemble and its surroundings is converged. Of course, we cannot disprove that our aimless shooting simulations avoid sampling unidentified zones of the transition state ensemble, separated from the ones we sample by barriers in trajectory path space. 

Finally, for each material and droplet size, among the accepted shooting point configurations, we randomly select between 110 and 210 structures for which we compute $P_{\text{crys}}$, this time using 50 independent unbiased dynamics instead of 10. Altogether with this procedure, we sample configurations from the transition state ensemble, which we define as $0.3\leq P_{\text{crys}} \leq 0.7$. In average, for each droplet size and materials, we obtain 90 structures matching this criterion and use them for the following structural analysis. Committor distributions are reported in the Supplementary Material. As one can see, most of the distributions are unimodal and peaked close to $P_{\text{crys}} = 0.5$, which indicates that we indeed sample configurations close to the true separatrix of the transition investigated. For some cases -- in particular the $N=1000$ and $N=2000$ Ag nanoparticles --, this is less evident, which can be due to insufficient sampling. In fact, since these distributions are skewed towards the crystalline state, the initial configurations selected for aimless shooting could have been too close to this state too, impacting the resulting committor distributions. Such effect could be alleviated with increased sampling. 

\subsection{Structural features of the transition states}

\begin{figure*}[ht]
    \centering
    \includegraphics[width=17cm]{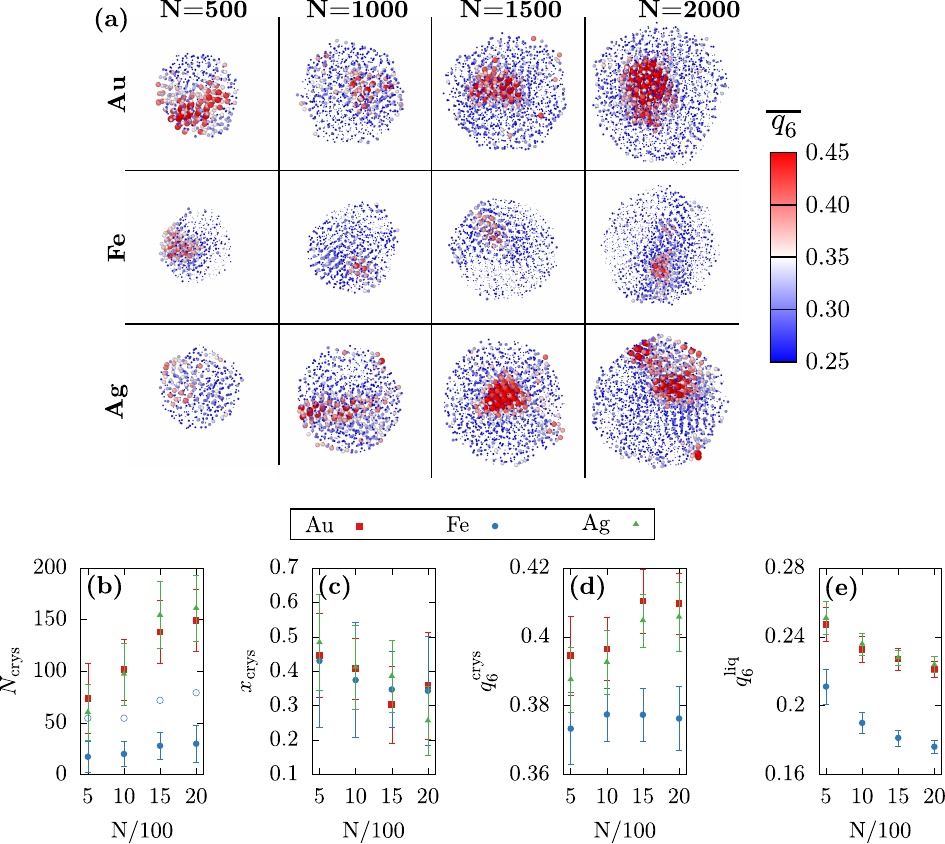}
    \caption{(a) For each droplet sizes and material types, one typical image of the critical structures is shown. Coloring from blue to red and atomic radius from small to large correspond to an increase in the value of $\overline{q_6}$. (b-e) Influence of the droplet size on the structure of the critical structures in terms of the size of the largest crystalline cluster (b), the  radial position of the  largest crystalline cluster (c), the value of $\overline{q_6}$ averaged over atoms of the largest crystalline cluster (d) and of the remaining atoms (e). We note that in (b) open circle correspond to $N_{\text{crys}}$ defined with a smaller threshold values for $\overline{q_6}>0.30$.}
    \label{Structure}
\end{figure*}

Fig.\,\ref{Structure} shows several measurements averaged over all of the obtained transition state configurations, of collective variables defined in Section \ref{sub:as}. As a preamble, we note that overall, all of the investigated structural features exhibit relatively large dispersion, which means that none of them should be considered as a collective variable optimally describing the crystallization process. First, one can observe that in panels (b)-(e), gold and silver seem to produce quantitatively similar results, which can be attributed to both systems having the same crystal structure (face centered cubic) and very similar lattice parameters (4.08\,\AA\, and 4.09\,\AA\, respectively for gold and silver). Meanwhile, for iron, values of $N_{\text{crys}}$ and $\overline{q_6}$ both in the liquid and the crystalline regime seem to be smaller than what is obtained for gold and silver. This can be explained by iron being intrinsically less ordered both in the crystalline and the liquid regime. In particular, since iron crystallizes in a body-centered cubic lattice, its overall $\overline{q_6}$ is deemed to be smaller\cite{lechner2008accurate} and if one defines a threshold of 0.30 instead of 0.35, we already see that $N_{\text{crys}}$ now gets much closer to values obtained with both gold and silver [See open points in Fig.\,\ref{Structure}]. After examining differences and similarities between the different materials, we will have a closer look at the influence of the nanoparticle size. First, $N_{\text{crys}}$ monotonically increases with larger nanoparticles [See Fig.\,\ref{Structure}.a]. Since we intentionally performed the simulations at the same degree of supercooling -- which would have led to similar critical nucleus sizes in bulk --, this result is far from being trivial and advocates for the crucial role of surfaces and finite size effects. Secondly, we observe that the value of $x_{\text{crys}}$ slightly yet consistently increases when going towards smaller nanoparticles. This means that the nuclei seem to get closer to the surface, favoring a somewhat heterogeneous nucleation mechanism, as opposed to homogeneous nucleation within the core of the nanoparticle as obserbed in the largest droplet sizes. In Fig.\,\ref{Structure}(d,e), we focus on the ordering quality of the critical nucleus ($q_6^{\text{crys}}$) and of the rest of the nanoparticle ($q_6^{\text{liq}}$). When decreasing the nanoparticle size, $q_6^{\text{crys}}$ and $q_6^{\text{liq}}$ have opposite trends, with the critical nucleus showing reduced crystallinity, and the surrounding liquid domain getting more structured. Compared to the evolution of $N_{\text{crys}}$, it seems that the decrease in the size of the crystalline cluster is compensated by a pre-ordered surrounding liquid phase. Observation of such a pre-ordered phase goes against classical nucleation theory and has already been observed in the bulk solidification of nickel and of molybdenum, using transition path sampling techniques\cite{diaz2018maximum,Menon2020Sep}. 

\begin{figure}[h!]
    \centering
    \includegraphics[width=8.6cm]{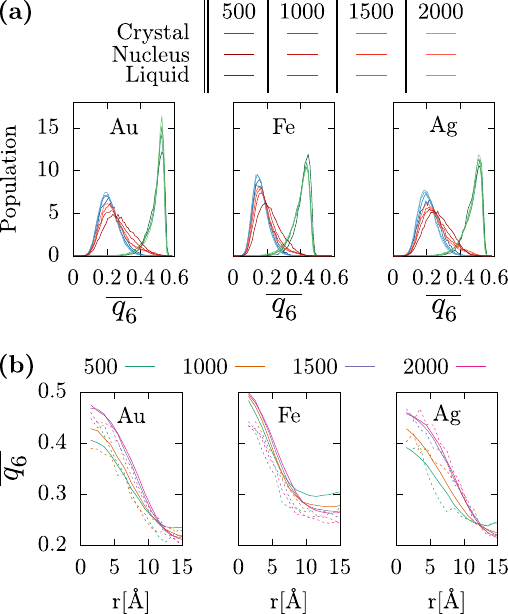}
    \caption{(a) Distribution of $\overline{q_6}$ measured for each droplet sizes and each materials when the whole nanoparticle is in the crystalline, critical and liquid regimes respectively in green, red and blue. (b) Spatial distribution of $\overline{q_6}$ measured with respect to the center of the critical nucleus for each droplet sizes and each materials and averaged over all of the obtained transition states. In dashed line, we plot one example of the same distribution but for a single transition state to show that the gradient is found in all cases.}
    \label{Prenuc}
\end{figure}

In order to further investigate this striking result, we display in Fig.\,\ref{Prenuc} two kinds of $\overline{q_6}$ distributions. Fig.\,\ref{Prenuc}.a compares the $\overline{q_6}$ distribution in the whole nanoparticle depending if the system is in the liquid, crystalline or critical state. It appears that while the crystalline and liquid regimes are not affected by changing $N$, the critical nucleus exhibits a shift in the distribution when decreasing the nanoparticle size, which is consistent with our previous findings. More importantly, we also observe a broadening of the peak, which suggests that nucleation for the smallest droplet requires a more diverse structural landscape. In addition, in Fig.\,\ref{Prenuc}.b, one can see the spatial distribution of $\overline{q_6}$ measured with respect to the center of the critical nucleus. First, the plateau at distances far from the critical nucleus, where atoms are mostly in the liquid state, is again consistent with our previous findings since we observe a higher value for the smallest nanoparticles, particularly clear for $N=500$. More importantly, while classical nucleation theory predicts a step-like behavior with the system being either in the liquid or the crystalline regime, a more complex scenario seems to emerge here, with a decreasing gradient in ordering  from the center of the crystalline nucleus. Increasing the nanoparticle size leads to a steeper gradient, with a less diffuse interface between the crystalline nucleus and the rest of the droplet. It is therefore clear that spatial confinement affects the crystallization mechanism, with greater departure from classical nucleation theory with stronger nanoscale effects. 

Altogether, our results advocate for a departure from classical nucleation theory at nanoscale, including diffuse interfaces between the critical cluster and the rest of the structure, as well as the existence of a pre-ordered liquid phase favoring the emergence of crystals, which is in-line with previous results obtained so far in bulk systems\cite{diaz2018maximum}. Here, we show that such behavior is enhanced by surface and finite-size effects inherently present at the nanoscale.

% https://journals.aps.org/prl/abstract/10.1103/PhysRevLett.94.235703
% papier peut-être intéressant pour comparer au bulk et/ou interpréter la variabilité de structure/cristallinité du noyau critique, même si c'est du LJ ici

\section{Conclusions}

Using numerical simulations in the form of molecular dynamics, we have investigated the crystallization mechanism of gold, silver, and iron nanoparticles of various sizes, with a focus on structurally analyzing  configurations located at the transition state between the crystalline and liquid states. For each composition, after determining the melting temperature as a function of the nanoparticle size, we perform simulations in the supercooled regime. Initial reactive trajectories are obtained using biased simulations in the form of metadynamics, from which we coarsely identify transition states and pathways, using unbiased committor analysis. We then obtain many brute-force reactive trajectories using transition path sampling in the form of aimless shooting, from which we extract representative transition state structures. An analysis of these structures reveals several systematic trends that are directly related to the reduction in size down to the nanoscale, \textit{i.e.} surface and finite-size effects: with smaller nanoparticle sizes, the critical nucleus gets smaller and less ordered, and is shifted towards the surface of the nanoparticle. In the meantime, the surrounding atoms show increased ordering, with the emergence of a pre-ordered liquid phase for the smallest nanoparticles. Finally, with smaller nanoparticle sizes, it is less evident to differentiate between the critical nucleus and the rest of the system, as the interface between both phases gets more diffuse, and both phases are less contrasted. Overall, this indicates a departure from classical nucleation theory induced by nanoscale, which hints at the importance of characterizing explicitly crystallization kinetics and thermodynamics for nanoparticles without resorting to classical models relying on assumptions on transformation mechanisms. 

Future directions would include analyzing the entire transformation pathway from the liquid to the crystalline phase, a complementary analysis to the present one focusing on the transition state ensemble. Ultimately, we should be able to compare different collective variables \textit{via} their ability to correlate with the committor, not only at the top of the free energy barrier, but also along the entire reaction process. Such an analysis, combined with the estimation of the probability that a trajectory passing through a certain configuration connects both basin, would allow us to determine which key features characterizing the pre-ordered liquid leads to a reactive path. In closing, we expect that our original approach can easily be transferred towards different nanoscale systems including oxydes and nanoalloys to unravel complex mechanisns associated with polymorphs stability and chemical ordering.

%
%
%\textbf{Structure} (fig3) [viz: or, cv: or, argent, 2LJ]
%
%\begin{itemize}
%    \item $N_{\text{cry}}$: augmente avec N. Plus la nanoparticule est grande, plus le noyau cristallin est grand. Cela prouve qu'il y a des effets de surface. 
%    \item $N_{\text{cry}}/N$ (proportion cristalline): reste relativement constante avec $N$ (aux barres d'erreur près). Peut-être qu'elle réduit un peu de manière systématique pour Ag. 
%    \item $q_6^{\text{cry}}$: augmente avec $N$. Plus la nanoparticule est grande, plus le noyau critique est structuré. Moins d'effets de surface peut-être?
%    \item $q_6^{\text{liq}}$: diminue avec $N$. La fraction non-cristalline à l'état de transition est moins structurée. On se rapproche d'un comportement "bulk"? Ici il est très important de savoir exactement comment on définit la fraction cristalline ($q_6>0.25?$)
%    \item $x_{\text{cry}}$: Noyau excentré pour Au à $N=500$, sinon plus ou moins centré (expliquer différence avec Ag? $N_{\text{cry}}(\text{Ag}) > N_{\text{cry}}(\text{Au})$ donc plus "centrable"?)

%\end{itemize}

\section*{Supplementary Material}
The supplementary material includes three sections: 1) Disentangling temperature and droplet size effects, 2) Aimless shooting additional figures and 3) Influence of \textit{ad hoc} parameters.

\section*{Acknowledgment}

This study was supported by the French National Research Agency (ANR) in the framework of its “Jeunes chercheuses et jeunes chercheurs” program, ANR-21-CE09-0006 and by the EUR grant NanoX n° ANR-17-EURE-0009 in the framework of the "Programme des Investissements d’Avenir". Computational resources have been provided by CALMIP, by Jean Zay and by TGCC.

\end{document}